\documentstyle[twoside,fleqn,espcrc2,epsf]{article}

\newcommand{\vev}[1]{\langle #1 \rangle}

\newcommand{\be}{\begin{equation}}
\newcommand{\ee}{\end{equation}}

\newcommand{\AmS}{{\protect\the\textfont2
  A\kern-.1667em\lower.5ex\hbox{M}\kern-.125emS}}

\title{Semileptonic B Decays from an NRQCD/D234 Action}

\author{J.Shigemitsu\address{Physics Department, The Ohio State
        University, Columbus, OH 43210, USA.},
        S.Collins$^{\rm b}$, C.T.H.Davies\address{Department of Physics \&
               Astronomy, University of Glasgow, Glasgow, G12 8QQ, UK.},
        J.Hein\address{Department of Physics \& Astronomy, University 
                 of Edinburgh, Edinburgh, EH9 3JZ, UK.},
       R.R.Horgan\address{D.A.M.T.P., CMS, Wilberforce Rd., Cambridge,
                England CB3 0WA, UK.},
       G.P.Lepage\address{Newman Laboratory, Cornell University, 
                 Ithaca, NY 14853, USA.}
              }

\begin{document}

\begin{abstract}
Semileptonic $B$ decays are studied on quenched anisotropic lattices 
using Symanzik improved glue, NRQCD heavy quark and D234 light quark 
actions.  We employ constrained fits to extract ground state contributions 
to two-  and three-point correlators.  Results for the $B \rightarrow 
\pi, l \overline{\nu}$ decay form factors are compared with previous 
lattice results. We find that our systematic errors (excluding 
quenching errors) are dominated by chiral extrapolation uncertainties.
\vspace{1pc}
\end{abstract}

\maketitle

\section{Introduction}
Several quenched lattice studies of $B \rightarrow \pi, l \overline{\nu}$ 
semileptonic decays have been completed in recent years \cite{fermilab}.
 They represent the first steps towards providing 
experimentalists and phenomenologists with the hadronic matrix 
elements involving the $b$ quark
that are  necessary to carry out precision consistency tests of the 
Standard Model.  In this article we present further studies of 
form factor calculations in the quenched approximation.  We work with 
glue and quark actions that are more highly improved than in 
previous lattice calculations,  we employ anisotropic lattices to 
enhance signal to noise and use constained fits \cite{bayes} 
 to extract groundstate 
contributions to correlators in a controlled way.  Simulation details,
including the precise form of the action employed, can be found 
 in \cite{semil}. We work at 5 values of the light quark mass spanning
$0.7 \, m_{strange} \leq  m_q \leq 1.3 \, m_{strange}$ and at one value 
of the heavy quark mass tuned to the $b$ quark mass using the $B_s$ meson. 
 Results are 
then extrapolated to the chiral limit.

\section{ Fitting Three-point Correlators}
The starting point for a form factor calculation on the lattice is 
the three-point correlator,
\begin{eqnarray}
\label{threepnt1}
  && C_\mu^{(3)}(\vec{p}_B,\vec{p}_\pi,t_B,t) = \sum_{\vec{x}}\sum_{\vec{y}}
e^{-i\vec{p}_B \cdot\vec{x}} e^{i(\vec{p}_B - \vec{p}_\pi)\cdot \vec{y} }
 \nonumber \\
  & &\vev{0|\Phi_{B}(t_B,\vec{x})\,V^L_\mu(t,\vec{y})\, \Phi^\dagger_\pi(0)
|0} .
\end{eqnarray}
$\Phi_\pi^\dagger$ and $\Phi_B^\dagger$ are interpolating operators 
used to create the pion or $B$ meson respectively.  $V^L_\mu$ is the 
dimensionless Euclidean space 
lattice heavy-light vector current. 
The continuum Minkowski space $V_\mu$ is related to $V^L_\mu$ via
\be
V_\mu = a_s^{-3} Z_{V_\mu}\,\xi_\mu\, V^L_\mu .
\ee
$\xi_\mu$ is the conversion factor between Minkowski and Euclidean 
space $\gamma$-matrices and $Z_{V_\mu}$ is the heavy-light current 
matching coefficient.  We work with $B$ mesons decaying at rest ($\vec{p}_B
=0$). $t_B$ is held fixed and we 
vary $t$, the location of the current insertions. 
The quantity $\xi_\mu C^{(3)}_\mu$ is fit, using constrained (Bayesian)
 fits, to the form,
\be
\label{threepnt}
\xi_\mu \,C^{(3)}_\mu(p_\pi,t)  = 
   \sum_j^{N_B} \sum_l^{N_\pi}
 A_{jl}e^{-E_\pi^{(l)}  t} e^{-E_B^{(j)}(t_B-t)}
\ee
Constrained fits allow for multi-exponential fits to a single 
correlator without loosing stability.
One augments the conventional $\chi^2$ with a term, $\chi^2_{prior}$, which 
prevents fit parameters that are not constrained by the data from 
taking on ``unreasonable'' unphysical values and thus destabilizing the fit.
\be
\label{chiaug}
\chi^2 \longrightarrow \chi^2_{aug} \equiv \chi^2 + \chi^2_{prior},
\ee
with
\be
\chi^2_{prior} \equiv \sum_j  \frac{(\alpha_j - \tilde{\alpha}_j)^2}
{\tilde{\sigma}^2_j} .
\ee
One can check that those parameters, such as groundstate energies and 
amplitudes, that are well determined by the data, are only minimally 
affected by $\chi^2_{prior}$. In Figs. 1 and 2 we show fit results for 
the groundstate amplitude $A_{11}$  for $\vev{V_0}$ 
 as a function of the 
number of exponentials $N_\pi$. $N_B$ is held fixed at 1 or 2 and 
fits are carried out to all data points between $t_{min}=1$ and 
$t_{max}=17 \sim 22$.  The numbers below data points give the 
$\chi^2_{aug}/d.o.f.$ of the fits. 
 One sees that fit results, including errors, 
stabilize once $N_\pi > 2$.  The fancy stars give bootstrap results 
at fixed $N_\pi = 3$ which we used in our final analysis (see 
references \cite{bayes,semil} for a description of constrained bootstrap fits).

\begin{figure}
\epsfxsize=7.0cm
\centerline{\epsfbox{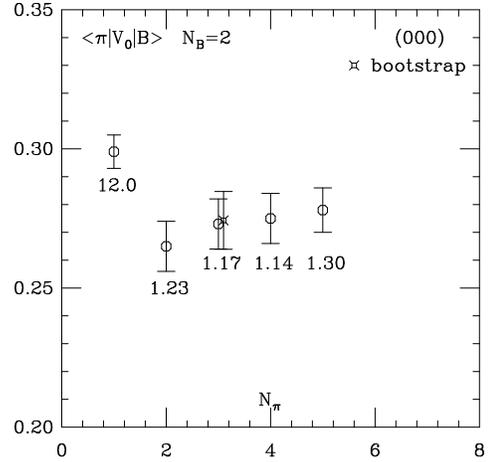}}
\caption{Fit results for the groundstate amplitude $A_{11}$ from 
the $\langle V_0 \rangle$ threepoint correlator. The numbers below 
the data points give $\chi^2_{aug}/d.o.f$.  The pion momentum is zero.
 }
\end{figure}

\begin{figure}
\epsfxsize=7.0cm
\centerline{\epsfbox{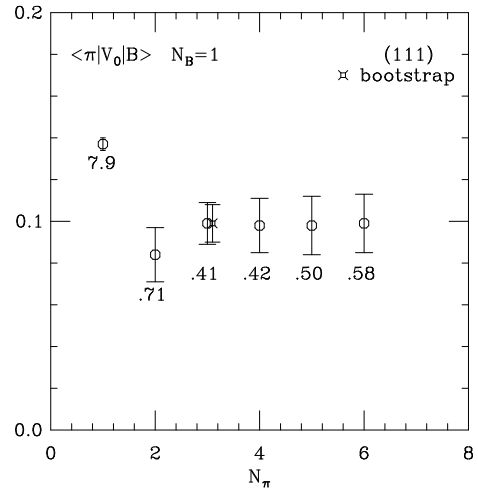}}
\caption{ Same as Fig. 1 for pion momentum (111) in units 
of $2\pi/(a_sL)$.
 }
\end{figure}

\section{Results for Form Factors and Comparisons with Previous Work}
The groundstate amplitudes $A_{11}(V^L_\mu)$ extracted from
three-point correlators  are straightforwardly related to the 
continuum matrix element of interest.
\be
\label{a11}
 \langle \pi(p_\pi) | \, V^\mu \, | B\rangle
 =
 \frac{A_{11}(V^L_\mu)}
{\sqrt{\xi^{(1)}_\pi \xi^{(1)}_B}} \,2 \, \sqrt{E_\pi M_B} \, Z_{V_\mu}
\ee
$\xi^{(1)}_\pi$ and $\xi^{(1)}_B$ are fixed from $\pi-\pi$ and $B-B$ 
 two-point correlators.
Fig. 3 shows results for the form factors $f_+(q^2)$ and $f_0(q^2)$ 
with the light quark mass around the strange mass.  Errors reflect 
only statistical errors.  Fig. 4 shows results after chiral extrapolation 
to the physical final state pion.  One sees that errors have increased 
considerably.

\begin{figure}
\epsfxsize=7.0cm
\centerline{\epsfbox{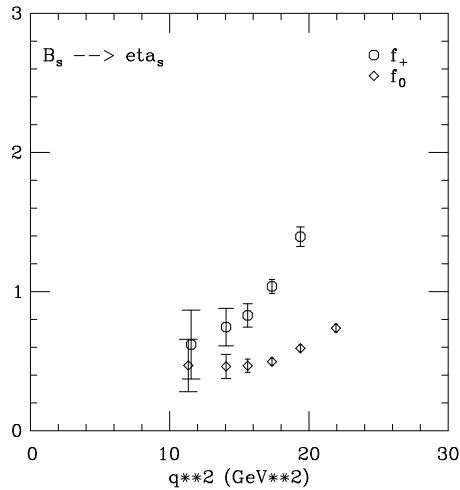}}
\caption{The form factors $f_+$ and $f_0$ for the light quark mass fixed 
at the strange quark mass.  Only statistical errors are shown.
  }
\end{figure}

\begin{figure}
\epsfxsize=7.0cm
\centerline{\epsfbox{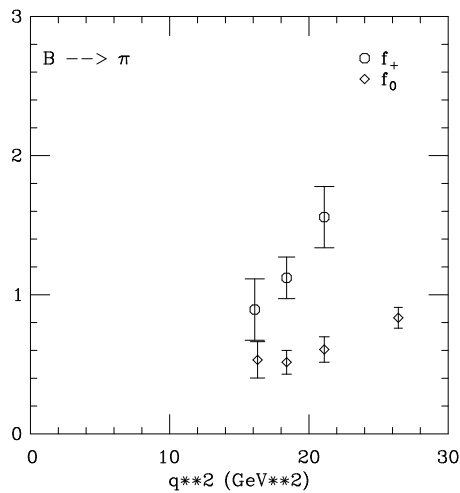}}
\caption{The form factors $f_+$ and $f_0$ after chiral extrapolation 
to the physical pion.
  Statistical and chiral extrapolation errors shown.
 }
\end{figure}
In Fig. 5 we compare our results with those of other collaborations.  Despite 
large differences in sources of systematic errors, one sees general 
agreement between  all groups, especially for the form factor 
$f_+$.  All calculations suffer from large errors. 
In most cases, in particular in the present study,
 systematic errors are dominated 
by the chiral extrapolation uncertainties.

\begin{figure}
\epsfxsize=7.0cm
\centerline{\epsfbox{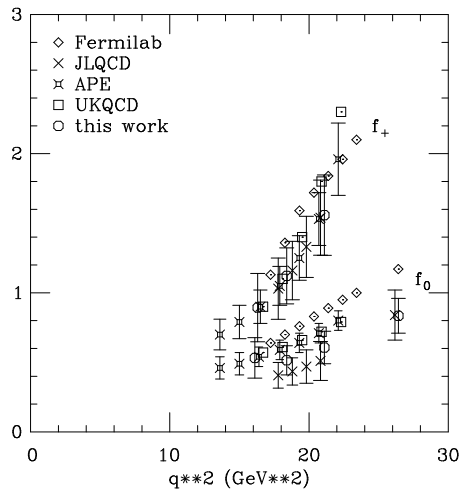}}
\caption{Comparison of form factors from different lattice groups. 
Some errors are omitted.
 }
\end{figure}

\vspace{.1in}
\noindent
Acknowledgements : This work was supported by DOE Grant
 DE-FG02-91ER40690 and by PPARC and NSF.

\end{document}